# Structure and Stability of Two Dimensional Phosphorene with =O or =NH Functionalization

Jun Dai[a] and Xiao Cheng Zeng[a*]

**We investigate stability and electronic properties of oxy- (=O) or imine- (=NH) functionalized monolayer phosphorene with either single-side or double-side functionalization based on density-functional theory calculations. Our thermodynamic analysis shows that oxy-functionalized phosphorene can be formed under the conditions ranging from ultrahigh vacuum to high concentrations of molecular $O_2$, while the imide-functionalized phosphorene can be formed at relatively high concentration of molecular $N_2H_2$. In addition, our Born-Oppenheimer molecular dynamics (BOMD) simulation shows that at the ambient condition both $O_2$ and $N_2H_2$ can etch phosphorene away.**

Recently, a new two-dimensional (2D) semiconductor, phosphorene, has been successfully isolated through mechanical cleavage of crystalline black phosphorus.[1-4] It has been demonstrated that phosphorene-based field-effect transistors (FETs) can exhibit high on/off ratio (~$10^5$) and relatively high carrier mobility (up to 1000 cm$^2$/V/s),[1, 2, 4] suggesting potential applications of phosphorene in nano-electronic devices. Black phosphorus, the bulk counterpart of phosphorene, is the most stable form of phosphorus and was discovered by Bridgman in 1914.[5] Like graphite, black phosphorus is also a layered material with weak interlayer van der Waals (vdW) interaction. In each layer, the phosphorus atom is bonded with three adjacent phosphorus atoms, forming a puckered honeycomb structure.[6, 7] The three bonds take up all three valance electrons of phosphorus atom. Thus, a monolayer phosphorene is a 2D semiconductor with a direct bandgap of about 0.3 eV.[8-12] More interestingly, the direct bandgap feature of few-layer phosphorene is dependent on the thickness. Previous theoretical calculations show that the bandgap can be tuned from 1.5 eV for monolayer to 0.6 eV for 5-layer phosphorene.[13] Furthermore, either in-plane or out-of-plane strain can significantly change the bandgap of monolayer phosphorene. Indeed, a ~5% in-plane strain can convert the monolayer phosphorene from a direct-gap to an indirect-gap semiconductor,[2] while a vertical compression can induce the semiconductor-to-metal transition.[14] The in-plane anisotropic optical properties of phosphorene[15, 16] and its potential applications in solar-cell systems[17] have also been reported.

Although black phosphorus is the most stable allotrope of phosphorus, it is still reactive under ambient condition.[5, 18-20] Especially, it has been reported that phosphorene flakes can be etched away at ambient condition.[20] We also note that for graphene, chemical functionalization can be an effective way to tune its electronic properties.[21-29] Therefore, it is useful to study stability and properties of chemical functionalized phosphorene sheets. Note also that carbon and phosphorus have different valence electron configuration, namely, $2s^22p^2$ for carbon and $3s^23p^3$ for phosphorus. In graphene, three valence electrons in carbon form $sp^2$ orbitals and the remaining valence electrons, one in each carbon atom, form delocalized π orbital. On the other hand, in phosphorene, phosphorus forms $sp^3$ bonding with a lone pair of valence electrons in each phosphorus atom. Therefore, the chemical species used for chemical functionalization of graphene may not be suitable for functionalization of phosphorene.

Here, we consider divalent electron donors such as =O, =S, =NH and =$CH_2$ for possible phosphorene functionalization in view of successful synthesis of organophosphonates whose simplest forms include O=$PH_3$, S=$PH_3$, HN=$PH_3$ and $H_2$C=$PH_3$.[30, 31] The structures and thermodynamic stabilities of these divalent ligands functionalized phosphorene are carefully examined. Effect of the functionalization on the electronic structure of phosphorene is also discussed. We show that oxy- (=O) functionalized phosphorene can be automatically formed in the presence of $O_2$ with either high or low $O_2$ concentration, while imide- (=NH) functionalized phosphorene can be automatically form in relatively high concentration of $N_2H_2$. Our BOMD simulations show that both oxy- and imide-functionalization can etch phosphorene away at the ambient condition in the presence of either $O_2$ or $N_2H_2$.

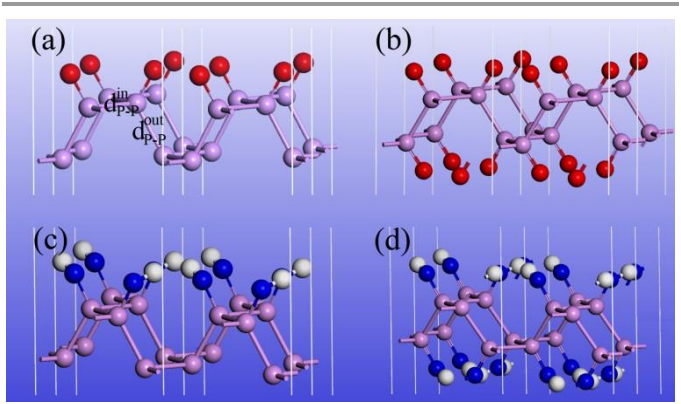

**Fig. 1** Optimized structure of (a) P-O-half, (b) P-O, (c) P-NH-half and (d) P-NH. Red, white, blue and pink spheres denote oxygen, hydrogen, nitrogen and phosphorus atoms, respectively. $d_{P-P}^{out}$ and $d_{P-P}^{in}$ denotes the P-P bond length in and out of the xy plane.

For density-functional theory (DFT) calculations, the generalized gradient approximation (GGA) for the exchange-correlation potential is adopted. The plane-wave cutoff energy for wave function is set to 500 eV. The ion-electron interaction is treated with the projected augmented wave (PAW)[32, 33] method as implemented in the Vienna *ab initio* simulation package (VASP 5.3).[34, 35] For geometry optimization, a 8×10×1 Monkhorst-Pack *k*-mesh is adopted for functionalized phosphorene systems. A vacuum spacing of ~20 Å is used so that the interaction between adjacent layers can be neglected. During the geometric optimization, both lattice constants and atomic positions are relaxed until the residual force on atoms are less than 0.01 eV/Å and the total energy change is less than $1.0\times10^{-5}$ eV. In addition, a combination of optB88-vdW[36, 37] for geometry optimization and HSE06[38] for band structure calculation (based on the optB88-vdW optimized structure) is used, which has been proven very reliable for few-layer phosphorene systems.[2] Our benchmark calculations for bulk black phosphorus also confirm reliability of the selected two DFT methods (see ESI Fig. S1).

Four divalent adsorbates, namely, =O, =S, =NH and =CH$_2$ are initially considered for functionalization of the phosphorene monolayer. However, our calculations show that S atoms cannot effectively bond to P atoms while =CH$_2$ can disrupt the integrity of phosphorene (see ESI Fig. S2). Hereafter, we only focus on =O and =NH functionalized phosphorene. Both single-side and double-side functionalization are taken into account. For simplicity, we use P-O-half and P-O notations to denote single-side and double-side =O functionalized phosphorene, and P-NH-half and P-NH for the single-side and double side =NH functionalized phosphorene. The optimized structures are shown in Fig. 1 and the structural parameters of P-O-half, P-O, P-NH-half, and P-NH are summarized in Table 1. One can see that the functionalization with four different patterns on phosphorene share some common features. Firstly, the functionalization with =O and =NH results in an in-plane structural expansion, where the in-plane lattice constant *a* expands over 4.9%~12.5% while *b* expands over 11.7%~21.0% with different functionalization patterns. As a result, the in-plane P-P bond length also expands. The P-P bond length that is out of the *xy* plane expands over 1.3%~5.6%. These expansions stem from the weakening of bond strength since these adsorbates are electron acceptors which can fetch electrons from P atoms (see ESI Table S1). Another intriguing feature is that in P-O-half, the difference in bond length for two types of P-P bonds is almost negligible.

**Table 1.** Computed equilibrium geometry parameters, including P-P bond length in the *xy* plane ($d_{P-P}^{in}$), P-P bond length out of the *xy* plane ($d_{P-P}^{out}$), P-X (X=O, N) bond length ($d_{P-X}$), in-plane lattice constants (*latt. const.*), and adsorption energies for functionalized phosphorene monolayer. **P-O-half**, **P-O**, **P-NH-half**, **P-NH** and **P** denote monolayer phosphorene with single-side =O functionalization, double-side =O functionalization, single-side =NH functionalization, double-side –NH functionalization and pristine monolayer phosphorene, respectively.

| | $d_{P-P}^{in}$ (Å) | $d_{P-P}^{out}$ (Å) | $d_{P-X}$ (Å) | latt. const. (Å) | $E_{ad}$ (eV) |
|---|---|---|---|---|---|
| **P-O-half** | 2.283 | 2.283 | 1.478 | a=5.091, b=3.469 | 4.132 |
| **P-O** | 2.338 | 2.380 | 1.483 | a=5.514, b=3.690 | 4.184 |
| **P-NH-half** | 2.288 | 2.304 | 1.555 | a=5.136, b=3.492 | 2.634 |
| **P-NH** | 2.386 | 2.367 | 1.562 | a=5.466, b=3.719 | 2.754 |
| **P** | 2.220 | 2.253 | N/A | a=4.556, b=3.305 | N/A |

Adsorption energies for the functional groups are computed at the HSE06 level, which is defined as: $E_{ad} = -(E_{tot} - E_P - nE_X)/n$, where $E_{tot}$ is the total energy of the functionalized phosphorene with *n* X (X = O or NH) as ligands, $E_P$ is the energy of pristine phosphorene monolayer, and $E_X$ is the energy of an isolated X. According to this definition, a larger value of $E_{ad}$ means stronger adsorption. As shown in Table 1, the average adsorption energy in the double-side functionalization is slightly greater than that in the single-side functionalization. Besides, the adsorption of =O is significantly stronger than that of =NH, because =O is a stronger electron acceptor than =NH. Thus, the P-O bonding is stronger than P-N.

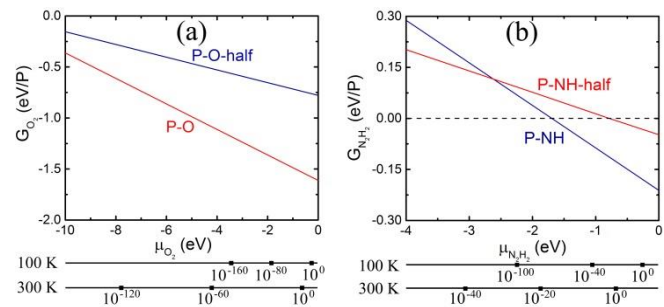

**Fig. 2** Formation energies versus chemical potential for (a) P-O-half and P-O, (b) P-NH-half and P-NH. The alternative axes show the pressure, in bar, of molecular O$_2$ and N$_2$H$_2$ gases corresponding to the chemical potential at T= 100 K and 300 K.

Similar to the definitions used in the discussion of the thermodynamic stability of functionalized graphene nanoribbons[39, 40] and graphene oxides[41], we define the zero

temperature formation energies of the functionalized phosphorene as :

$$E_f^X = \frac{1}{N}\left(E^{tot} - E_P - \frac{N_X}{2}E_X\right),$$

where $E^{tot}$ and $E_P$ are the total energy of the functionalized phosphorene and the pristine phosphorene, $E_X$ is the energy of the isolated $O_2$ or $N_2H_2$ molecule, $N_X$ is the number of O= or =NH in a supercell, and $N$ is the number of phosphorus atom in a supercell. The stability of different structures can be determined by $E_f^X$ under different experimental conditions. For example, in the presence of molecular $O_2$ gas, the relative stability can be obtained by comparing

$$G_{O_2} = E_f^{O_2} - \rho_O \frac{\mu_{O_2}}{2},$$

where $\rho_O = N_O/N$, at the absolute temperature $T$, and for a partial $O_2$ pressure $P$, the chemical potential of $O_2$ can be obtained as:

$$\mu_{O_2} = H°(T) - H°(0) - TS°(T) + k_B T \ln\frac{P}{P°},$$

where $H°(T)$ and $S°(T)$ are the enthalpy and entropy at the pressure $P°$ =1 bar (taken from the JANAF thermochemical tables).[42] For a given value of $\mu$, the structure with a lower value of $G$ is more stable.

In Fig. 2, $G$ vs $\mu$ is plotted for the functionalized phosphorenes. First, we can see that P-O is more stable than P-O-half for all negative values of the $O_2$ chemical potential, suggesting that the P-O is the more stable specie thermodynamically under experimental conditions ranging from ultrahigh vacuum to atmospheric concentration of molecular $O_2$. For P-NH and P-NH-half, we can see that P-NH-half is more stable than P-NH at relatively low values of $N_2H_2$ chemical potential, while at relatively high values of $N_2H_2$ chemical potential, P-NH is more stable than P-NH-half. Thus, indicates under the experimental condition of ultralow concentration of $N_2H_2$, P-NH-half is more stable, while under the condition of relatively high $N_2H_2$ concentration, P-NH is more stable than P-NH-half. Second, Fig. 2 suggests that phosphorene is unstable in the presence of $O_2$ even in the ultrahigh vacuum concentration of molecular $O_2$ ( $G$ is negative), and oxy-functionalized phosphorene can be automatically formed. In the presence of $N_2H_2$, phosphorene is unstable under condition of relatively high concentration of $N_2H_2$, and the imine-functionalized phosphorene can be automatically formed as well. To address the substrate induced lattice change in real systems, we computed the formation energies of P-O and P-NH with a uniaxial strain of 5% or -5% along $a$ or $b$ axis, or a biaxial strain of 5% or -5% along both $a$ and $b$. The results are summarized in ESI Table S2. One can see that a net effect of the strain on the G-μ curves as shown in Fig. 2 is a shift along $y$ axis in a range of +0.001 to +0.067 eV, therefore, will not change the conclusion regard the thermal stability of P-O and P-NH.

We have also studied thermal stability of the functionalized phosphorene by means of the BOMD simulations with the constant-pressure and constant-temperature (*NPT*) ensemble. Here, the pressure is set to 1 atm, while the temperature is controlled at either 70 K or 300 K. The time step is 2 fs, and the total simulation time is 8 ps. As shown in Fig. 3, at the low temperature (70 K), structures of P-O-half, P-O and P-NH-half are still robust and intact, indicating their stability near liquid nitrogen temperature. But for P-NH, the structure is partially destroyed, indicating at relatively high concentration of $N_2H_2$, phosphorene can be etched away. Near the room temperature (300 K), none of the four functionalized phosphorene sheets can retain their structure integrity, especially for the P-NH one which would decompose into several clusters with a huge volume expansion. The BOMD simulations indicate that these functionalized phosphorene sheets can be possibly observed at low temperature in the presence of either $O_2$ or $N_2H_2$, while at high temperature, phosphorene can be etched away.

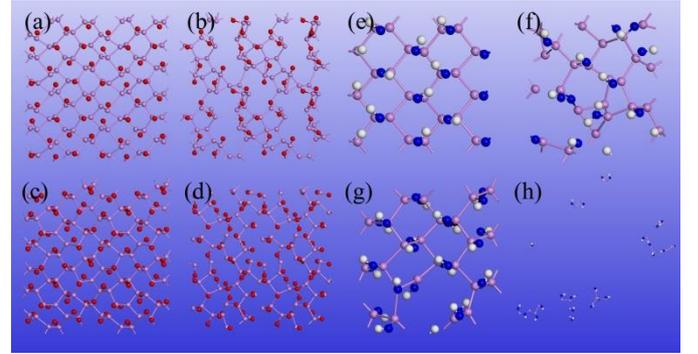

**Fig. 3** Snapshots of functionalized phosphorene monolayer at 8 ps of the Born-Oppenheimer molecular dynamics simulation in the NPT ensemble, (a) P-O-half at 70 K, (b) P-O-half at 300 K, (c) P-O at 70 K, (d) P-O at 300 K, (e) P-NH-half at 70 K, (f) P-NH-half at 300 K, (g) P-NH at 70 K and (h) P-NH at 300 K.

Lastly, we examine effects of the chemical functionalization on the electronic structures of phosphorene monolayer. Our benchmark calculations show that unlike half-hydrogenated or half-fluorinated graphene systems[43, 44] for which their magnetic properties can be tuned by surface functionalization, here the four functionalized phosphorene sheets appear to be non-spin-polarized. This is because phosphorus atoms in phosphorene adopt the $sp^3$ hybridization; the adsorbates act as electron acceptor and bond with P atoms with the electron lone pairs on P atoms. Hence, there is no unpaired electrons and spin polarization in these systems. The HSE06 band structures and atomic projected density of states (pDOS) are plotted in Fig. 4. Except for P-O, the direct bandgap feature is not retained in P-O-half, P-NH-half and P-NH. The bandgap of P-O-half (1.55 eV) is close to pristine monolayer phosphorene (~1.5 eV)[13]. The band gap in other three functionalized phosphorene sheets are slightly reduced to 1.03 eV, 1.44 eV and 1.24 eV for P-O, P-NH-half and P-NH, respectively. Moreover, in functionalized phosphorene sheets, the VBM is a hybrid state in which O or NH contributes the most part, while the CBM is nearly contributed by P atoms. It is worthy of mentioning that although previous theoretical calculations have shown that the

~5% in-plane compression can convert monolayer phosphorene from being a direct-gap to an indirect-gap semiconductor,[2] the studied phosphorene-based FET devices still have relatively good properties. Possible contamination of phosphorene by $O_2$ or $N_2H_2$, especially $O_2$, should also be carefully monitored since the contamination can harm performance of the devices.

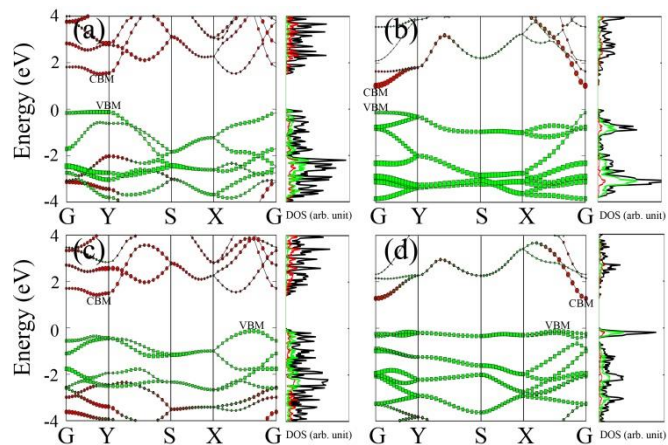

**Fig. 4** HSE06 band structure and atomic projected partial density of states (pDOS) of (a) P-O-half, (b) P-O, (c) P-NH-half and (d) P-NH. The Fermi level is set to 0. In band structures, the size of red spheres and green rectangles (lines) denote the contribution from P and O (or NH), and valence band maximum and conduction band minimum and labelled as VBM and CBM, while in pDOS, black, red and green lines denote the total DOS, pDOS of P and O (or NH), respectively.

## Conclusions

We investigate properties of oxy- (=O) and imine- (=NH) functionalized monolayer phosphorene with either single-side or double-side functionalization. Our thermodynamic analysis shows that in the presence of molecular $O_2$ with either high or low concentration, oxy-functionalized phosphorene will be automatically formed while imide-functionalized phosphorene will be formed at relatively high concentration of $N_2H_2$. Moreover, our BOMD simulation suggests that at ambient conditions the phosphorene will be etched away.

## Notes and references

[a]Department of Chemistry, University of Nebraska-Lincoln, Lincoln, NE 68588, USA. Email: xzeng1@unl.edu

# Electronic Supplementary Information

Structure and Stability of Two Dimensional Phosphorene with =O or =NH Functionalization

Jun Dai[1] and Xiao Cheng Zeng[1] [*]

[1]Department of Chemistry and Department of Mechanical and Materials Engineering, University of Nebraska-Lincoln, Lincoln, NE 68588, USA

Email: xzeng1@unl.edu

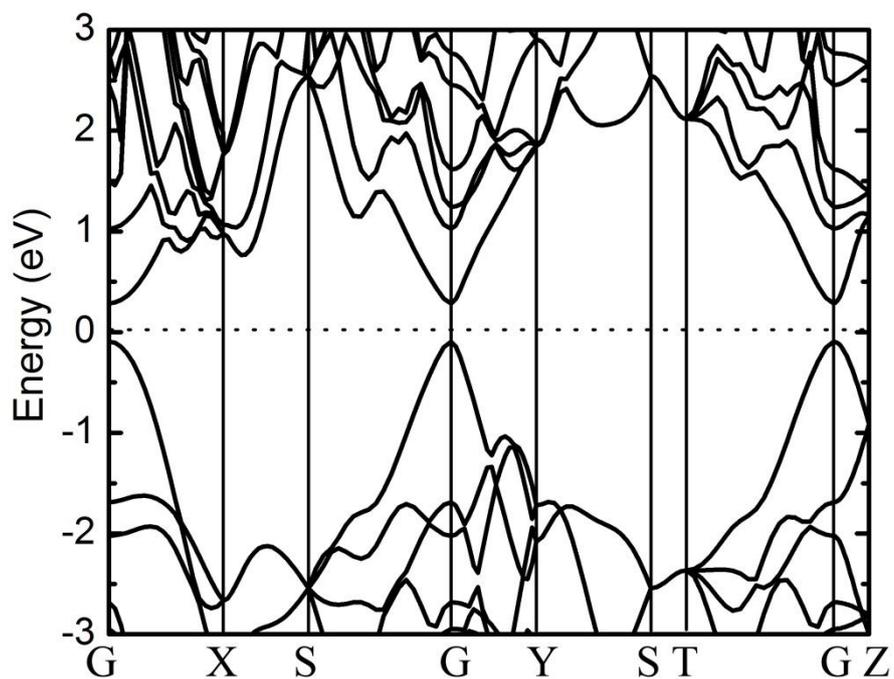

**Figure S1**. Computed band structures (HSE06) of bulk black phosphorus. G (0.0, 0.0, 0.0), X (0.0, 0.5, 0.0), S (0.5, 0.5, 0.00), Y (0.5, 0.0, 0.0), T (0.5, 0.5, 0.5) and Z (0.0, 0.0, 0.5) refer to special points in the first Brillouin zone.

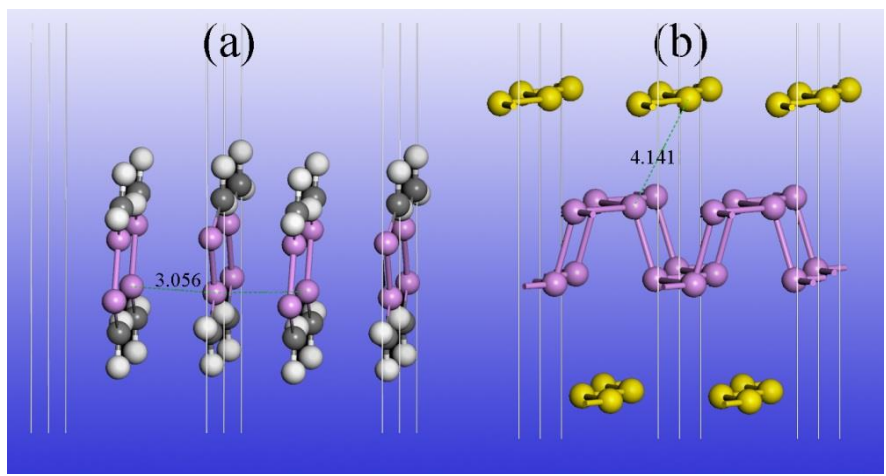

**Figure S2**. Optimized structures of =CH$_2$ and =S functionalized phosphorene, the pink, grey, white and yellow spheres denote the P, C, H and S atoms, respectively.

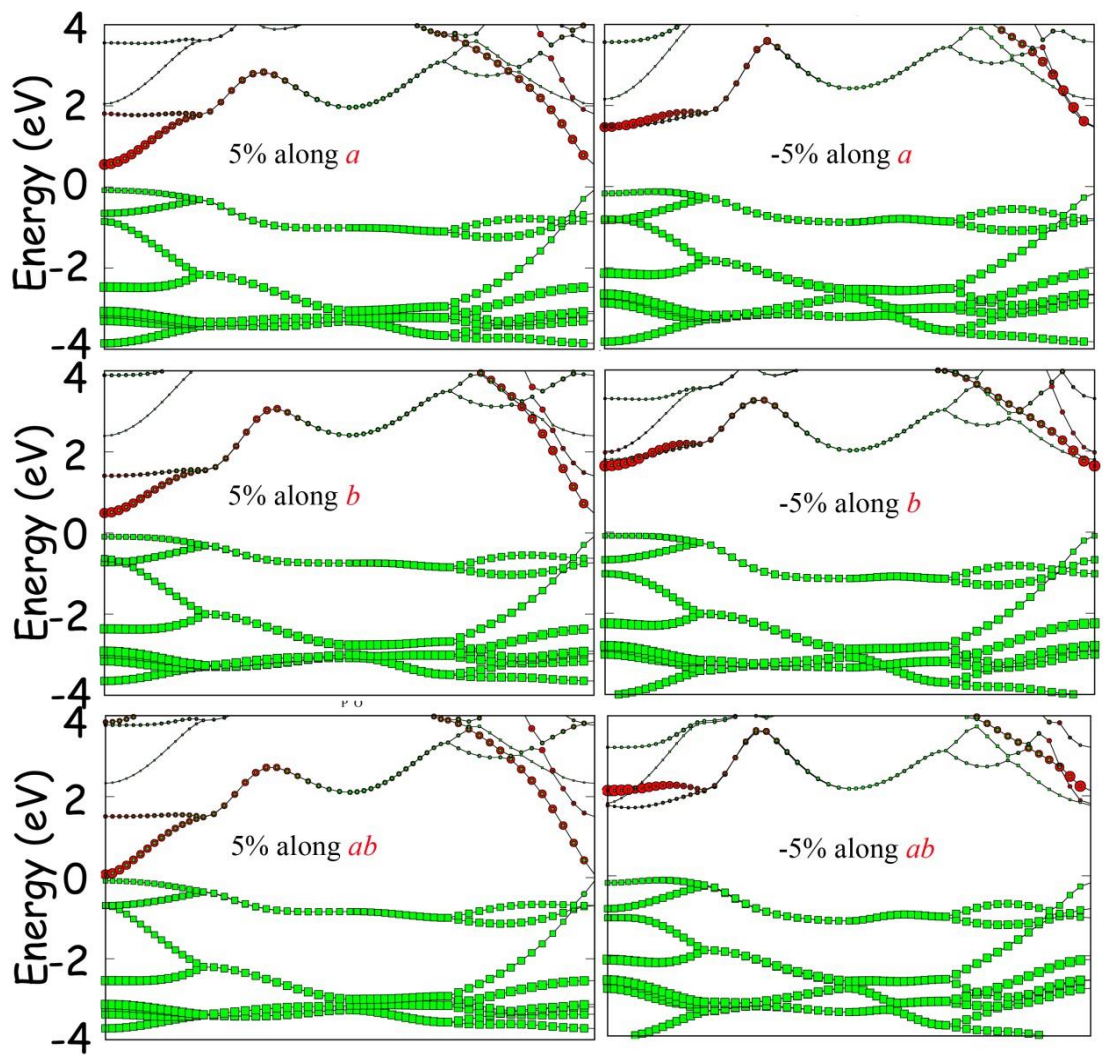

**Figure S3** HSE06 band structures of P-O under uniaxial or biaxial strain of 5% or -5%.

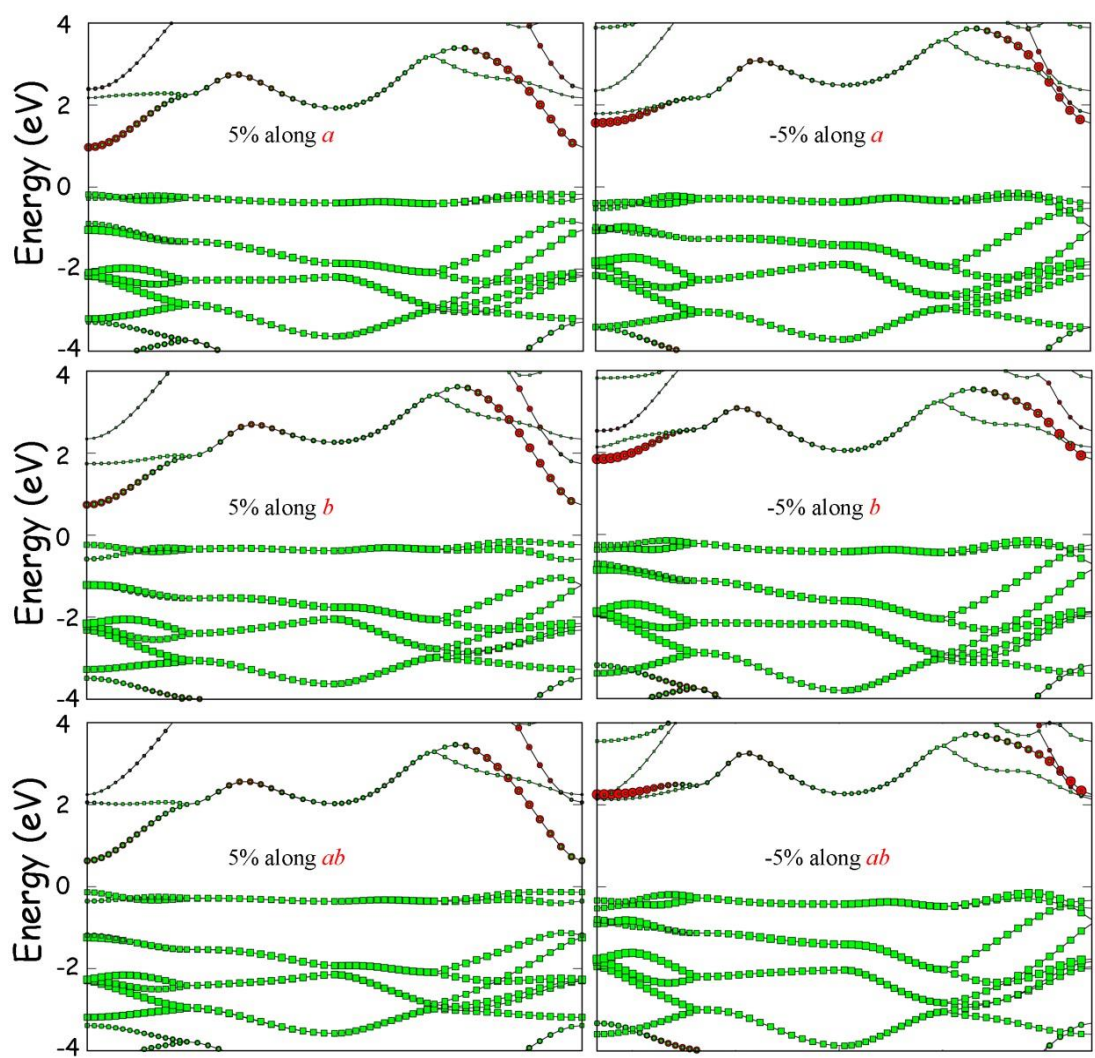

**Figure S4** HSE06 band structures of P-NH under uniaxial or biaxial strain of 5% or -5%.

**Table S1** Computed charge transfer for P-NH-half, P-NH, P-O-half and P-O based on Bader's charge analysis, $P^1$ denotes the phosphorus atoms with NH or O while $P^2$ denotes those without NH or O. The unit of the charge transfer is e⁻, positive values indicate electron gain while negative values electron loss.

|            | $P^1$  | $P^2$ | N     | H     | O     |
|------------|--------|-------|-------|-------|-------|
| P-NH-half  | -1.15  | +0.03 | +1.64 | -0.52 | N/A   |
| P-NH       | -1.12  | N/A   | +1.64 | -0.52 | N/A   |
| P-O-half   | -1.29  | +0.03 | N/A   | N/A   | +1.26 |
| P-O        | -1.27  | N/A   | N/A   | N/A   | +1.27 |

**Table S2** Computed formation energies (in eV/P) of P-O and P-NH with an uniaxial (along *a* or *b*) or a biaxial strains of +5% or -5%.

|  | +5% | 0% | -5% |
|---|---|---|---|
| P-O uniaxial along *a* | -1.588 | -1.612 | -1.604 |
| P-O uniaxial along *b* | -1.583 | -1.612 | -1.608 |
| P-O biaxial along *a* and *b* | -1.562 | -1.612 | -1.591 |
| P-NH uniaxial along *a* | -0.180 | -0.212 | -0.194 |
| P-NH uniaxial along *b* | -0.181 | -0.212 | -0.211 |
| P-NH biaxial along *a* and *b* | -0.145 | -0.212 | -0.195 |